\documentclass[prl,twocolumn,showpacs,superscriptaddress,amsmath]{revtex4-1}

\usepackage{amssymb}
\usepackage{graphicx}
\usepackage{dcolumn}
\usepackage{bm}
\usepackage[utf8]{inputenc}
\usepackage{mathtools}
\usepackage{verbatim}
\begin{document}
\title{Additional energy scale in SmB$_6$ at low temperature}

\author{L. Jiao}
\affiliation{Max-Planck-Institute for Chemical Physics of Solids,
N\"othnitzer Str. 40, 01187 Dresden, Germany}
\author{S. R\"o{\ss}ler}
\affiliation{Max-Planck-Institute for Chemical Physics of Solids,
N\"othnitzer Str. 40, 01187 Dresden, Germany}
\author{D.~J. Kim}
\affiliation{Department of Physics and Astronomy, University of California,
Irvine, CA 92697}
\author{L.~H. Tjeng}
\affiliation{Max-Planck-Institute for Chemical Physics of Solids,
N\"othnitzer Str. 40, 01187 Dresden, Germany}
\author{Z. Fisk}
\affiliation{Department of Physics and Astronomy, University of California,
Irvine, CA 92697}
\author{F. Steglich }
\affiliation{Max-Planck-Institute for Chemical Physics of Solids,
N\"othnitzer Str. 40, 01187 Dresden, Germany}
\author{S. Wirth}
\affiliation{Max-Planck-Institute for Chemical Physics of Solids,
N\"othnitzer Str. 40, 01187 Dresden, Germany}

\date{\today}
\maketitle
\textbf{Topological insulators give rise to exquisite electronic
properties due to their spin-momentum locked Dirac-cone-like band
structure. Recently, it has been suggested that the required
opposite parities between valence and conduction band along with
strong spin-orbit coupling can be realized in correlated materials.
Particularly, SmB$_6$ has been proposed as candidate material for a
topological Kondo insulator. By utilizing scanning tunneling microscopy
and spectroscopy measurements down to 0.35 K, we observed several states
within the hybridization gap of about $\pm$20 meV on well characterized
(001) surfaces of SmB$_6$. The spectroscopic response to impurities and
magnetic fields allows to distinguish between dominating bulk and surface
contributions to these states. The surface contributions develop
particularly strongly below about 7 K which can be understood in terms
of a breakdown of the Kondo effect at the surface. Our high-resolution
data provide detailed insight into the electronic structure of SmB$_6$,
which will reconcile many current discrepancies on this compound.}

\begin{figure}[tb]
\centering\includegraphics[width=6cm]{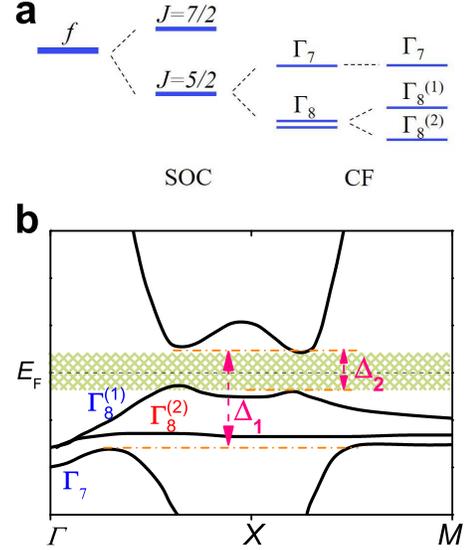}
\caption{\textbf{Sketch of the multiplet $f$-states and the resulting band
structure.} (\textbf{a}) Evolution of energy levels of the
$f$-states in SmB$_6$, which follows from the work of Ref.\
\cite{Baruselli,KangJPSJ}. The $f$-states are split into $J$ = 7/2 and
$J$ = 5/2 states by spin-orbit coupling (SOC). The $J$ = 5/2 state, which
is slightly below $E_F$, is split into a $\Gamma_7$ doublet and a $\Gamma_8$
quartet by the crystal field (CF). Away from the $\varGamma$ point, the
$\Gamma_8$ quartet is further split into $\Gamma_8^{(1)}$ and $\Gamma_8^{(2)}$
doublets. (\textbf{b}) A schematic bulk band structure of SmB$_6$ based on
calculations of Ref. \cite{Ale2013,Tak2011,Lu2013,Junwon,Baruselli,KangJPSJ}.
Hybridization between the $\Gamma_7$, $\Gamma_8^{(1)}$ bands and the
conduction band opens two gaps which are denoted as $\Delta_1$ (typically
around 20 meV) and $\Delta_2$. The shaded area marks the small bulk gap which
may host in-gap states. See also Supplementary Fig. 1 for details.}
\label{Fig0}
\end{figure}

In the past few years, the concept of strong topological insulators
which exhibit an odd number of surface Dirac modes characterized
by a $\mathbb{Z}_2$ topological index, has attracted great interest.
In this context, it was theoretically predicted that some Kondo
insulators, such as SmB$_6$, Ce$_3$Bi$_4$Pt$_3$, CeNiSn, CeRu$_4$Sn$_6$,
are candidates for strong three-dimensional (3D) topological insulators
\cite{Dze2010,Ale2013}. In particular, SmB$_6$ is intensively studied due
to its simple crystal structure and clear signatures of a hybridization
gap. Theoretically, a common picture of the multiplet $f$-states and the
Kondo coupling is shared among different band structure calculations
for bulk SmB$_6$ \cite{Ale2013,Tak2011,Lu2013,Junwon,Baruselli,KangJPSJ},
as sketched in Fig.\ \ref{Fig0}. Due to strong spin-orbit coupling and crystal
field effects, the $f$-states of Sm are split into several multiplets as
presented in Fig.\ \ref{Fig0}a. Considering the symmetry of the multiplets,
only the $\Gamma_7$ and $\Gamma_8^{(1)}$ bands are allowed to hybridize with
the Sm $d$-band via the Kondo effect \cite{Lu2013,Baruselli}. As a result, two hybridization gaps ($\Delta_1$, $\Delta_2$) may open at different energies as
sketched in Fig.\ \ref{Fig0}b (in principle only $\Delta_2$ is a well-defined
gap). Although topological surface states (TSS) are unambiguously predicted
to reside within the hybridization
gap \cite{Ale2013,Tak2011,Lu2013,Junwon,Baruselli,KangJPSJ}, no consensus has
been reached on the structure of the TSS around the Fermi energy ($E_F$).
Nonetheless, the prediction of TSS provides an attractive explanation
for the four decades-old conundrum \cite{Men1969} of SmB$_6$ which exhibits
a plateau in the resistivity typically below about 5 K \cite{All1979,Kim2014}.

Experimentally, the existence of metallic surface states below about 5 K has
been best illustrated by electrical transport measurements on SmB$_6$
\cite{Kim,Kim2014,Wol2013}. However, the \emph{origin} of these surface states
and their topological properties remain controversial in spite of intensive
investigations. Several properties of SmB$_6$ interfere with a straightforward interpretation. One major issue arises with respect to the size of the
hybridization gap. Spectroscopic measurements observed a large hybridization
gap of about 15--20 meV \cite{Zhang,Frantzeskakis,Den2014,Neupane,Jiang,Denlinger,Xu2013,Xu2014,Min2014,Xu,Zhu,Hlawenka},
which is normally understood by considering a single coherent $f$-band
hybridizing with a conduction band (Supplementary Fig.\ 1). Typically,
additional features within this energy scale are assumed to be in-gap
states. In some cases, the in-gap states are further ascribed to TSS
\cite{Neupane,Jiang}. On the other hand, analyses of thermal activation
energies derive a small excitation energy of 2--5 meV, which shows bulk
properties and is understood in terms of a small, likely indirect, bulk gap
\cite{Coo1995,Chen,Zhou16} or in-gap states \cite{Gab,Gor,Fla2001}.
Obviously, different probes as well as different ranges in the measurement
temperatures reveal only either the bigger or the smaller hybridization gap
sketched in Fig.\ \ref{Fig0}b. Nevertheless, these measurements provide
essential constraints to the sizes of the two hybridization gaps.
In terms of topology (\textit{i.e.} trivial or non-trivial surface states),
experimental results, even obtained by using the very same method, are
conflicting among many reports \cite{Den2014,Frantzeskakis,Neupane,Jiang,Denlinger,Xu2013,Xu2014,Min2014,Xu,
Zhu,Hlawenka,Li2014,Tan2015,Tho2013,Nakajima}.
Considering the exotic phenomena which appear only within $\pm$20 meV and
below 5 K, measurements with very high energy resolution and at very low
temperature are highly desired.

Another severe difficulty, which contributes to such a wide
discrepancy among the experimental results, is caused by the surface
itself. Specifically, the (001) surface of SmB$_6$ is polar
\cite{Zhu}. This can induce different types of band bendings
\cite{Den2014}, quantum well confinements \cite{Kan2015}, charge
puddles, and surface reconstructions \cite{Yee,Sahana,Ruan,roe15}.
Specifically the latter may give rise to conducting surface layers on its
own \cite{Zhu}. Frequently, different types of surfaces (B- and
Sm-terminated, reconstructed and non-reconstructed) coexist at different
length scales on one and the same cleaved surface which may complicate
interpretation of spectroscopic results, \textit{e.g.}, by angle-resolved
photoemission spectroscopy (ARPES).

We therefore conducted scanning tunneling microscopy/spectroscopy
(STM/STS) down to the base temperature of 0.35 K with an energy
resolution of about 0.5 meV. This allowed us to identify the fine
structure of the hybridization gaps on large and non-reconstructed
surfaces in the sub-meV scale. Moreover, by measuring the impurity,
magnetic-field and temperature dependence of the STS spectra, we
were able to attribute bulk and/or surface contributions to these
states, and unveil a new energy scale of $\simeq$ 7 K, which provides
an important piece of the puzzle for a unified picture of SmB$_6$.

\section*{Results}
\begin{figure*}[tb]
\centering\includegraphics[width=15cm]{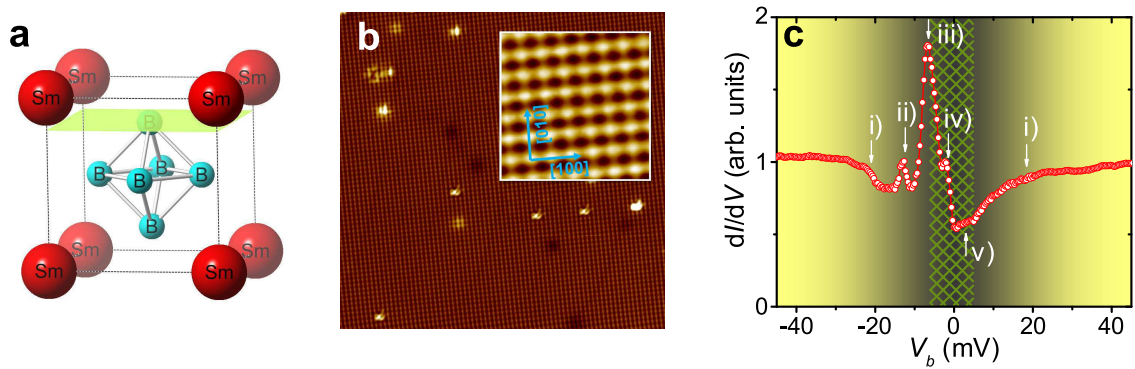}
\caption{\textbf{B-terminated surface and STS spectra at base temperature.}
(\textbf{a}) Cubic crystal structure of SmB$_6$ with lattice
constant $a$ = 4.133 \AA. The green plane indicates a cleave with
B-terminated surface. (\textbf{b}) STM topography on a $30\, \times\, 30$
nm$^2$ non-reconstructed B-terminated surface of SmB$_6$ ($T$ = 0.35
K, bias voltage $V_b$ = 300 mV, set-point current $I_{sp}$ = 400 pA).
Note the small number of defects. For the height scale compare to
Fig.\ \ref{Fig3}(\textbf{b}). The zoomed inset ($3\, \times\, 3$ nm$^2$)
shows the orientation of the crystallographic axis, clearly
indicating B termination. (\textbf{c}) Spatially ($2 \!\times\! 2$ nm$^2$)
averaged STS on part of the surface displayed in the inset of
(\textbf{b}). Several features can clearly be distinguished
within $\pm$20 mV, which are marked as i) to v) and discussed in
the text. Yellow to gray background in (\textbf{c}) indicates the
energy range within which the gap opens, while
the patterned area marks the region for potential in-gap
states. $V_b$ = 50 mV, $I_{sp}$ = 125 pA, modulation voltage
$V_{mod}$ = 0.2 mV.
} \label{Fig1} \end{figure*}

\textbf{Topography and STS spectra at base temperature.}
In Fig.\ \ref{Fig1}b, the topography of a non-reconstructed surface
with clear atomic resolution is presented. The distance of about 4.1
{\AA} and arrangement of the corrugations is in good agreement with
the lattice constant $a =$ 4.133 {\AA} of the cubic crystal structure
of SmB$_6$ (Fig.\ \ref{Fig1}a). The very small number of defects compared
to the number of unit cells within the field of view (more than 5200)
not only indicates high sample quality but also ensures that the measured
spectrum is not influenced by defects.
The absence of any corrugation other than along the main crystallographic
axes, as nicely seen in the inset to Fig.\ \ref{Fig1}b, clearly
indicates a B-terminated surface \cite{Sahana,roe15}.

The tunneling conductance $g(V)$ $\equiv$ d$I(V)/$d$V$, measured at
$T =$ 0.35 K and far away from any impurity, exhibits several anomalies
close to $E_{\rm F}$, marked by i) -- v) in Fig.\ \ref{Fig1}c.
A change in the slope of $g(V)$ around $\pm$20 meV suggests a
pronounced loss of local density of states (LDOS) within this energy
range. Around the same energy, the opening of a gap has been widely
observed by a number of spectroscopic tools as mentioned
above \cite{Frantzeskakis,Jiang,Denlinger,Xu2013,Xu2014,Min2014,Xu,
Zhu,Hlawenka}, including STS \cite{Yee,Sahana,Ruan}.
Based on the band structure displayed in Fig.\ \ref{Fig0}b, the kinks
marked by i) can be ascribed to the Kondo coupling between the $f$-band
and the conduction band, which results in a decreased conduction
electron density inside the hybridization gap below the Kondo temperature
$T_{K}$ \cite{ErnstNature2011}.

More importantly, we were able to disentangle several anomalies which were
hitherto not resolved \emph{individually} by STS at higher temperature
\cite{Yee,Sahana,Ruan}. Benefitting from this improvement, we can
investigate the fine structure of bulk/surface bands and go beyond a
simple Kondo hybridization analysis of the bulk states, which is based on
only one $f$-band and one conduction band \cite{Den2014}. Around $-13.5$
meV, there is a small peak marked by ii). Excitations with similar energy
have been reported before, {\it e.g.} by ARPES ($-15$ meV) \cite{Neupane},
X-ray photo\-electron spectroscopy (XPS) ($-15$ meV) \cite{Den2014} and
inelastic neutron scattering (14 meV) \cite{Fuhrman,ale95}, yet with
differing explanations as to its origin. As discussed below, this small
peak is most likely related to the indirect tunneling into the localized
$\Gamma^{(2)}_8$ states. Compared to delocalized $f$-states, such localized
$f$-states may give rise to only small anomalies in spectroscopy
measurements \cite{Ramankutty}.

Compared to peak ii), peak iii) (at around $-6.5$ meV) is very sharp and
pronounced. Such a peak has been observed on different types of
surfaces, including reconstructed ones \cite{Yee,Sahana,Ruan}, which clearly
indicates that there are significant bulk contributions to this state.
Very likely, the weakly dispersive structure of the hybridized
$\Gamma^{(1)}_8$ band around the $X$-point along with the Fano effect can
induce a peak in the conductance spectra at this energy level.
In a Kondo system, the Fano effect is due to a quantum mechanical
interference of electrons tunneling into the localized states and
the conduction bands \cite{mal09,Figgins}. Either a sharp drop (like
feature i)) or a pronounced peak will show up around the gap edge,
depending on the tunneling ratio between the two channels as well as the
particle-hole asymmetry of the conduction band. However, as has been
reported previously, the spectrum deviates from a simple Fano model at low
temperature \cite{Ruan,Yee}, indicating additional components to peak iii)
(see also discussion below). This is consistent with our inference that
the hybridized $\Gamma^{(1)}_8$ band resides within the big gap $\Delta_1$
and also contributes to the intensity of this peak. Hence, the position of
peak iii) can provide an indication with respect to the energy level of
the $\Gamma^{(1)}_8$ band and therefore the size of the small gap $\Delta_2$.
Note that its energy level is also comparable with the size of the small bulk
gap observed by transport measurements \cite{Coo1995,Chen,Zhou16}.
Therefore, peak i) to iii) can directly be compared to the band structure
in Fig.\ \ref{Fig0}b. To verify the bulk/surface origins of these peaks at
low temperature, impurity, magnetic field, and temperature dependence of STS
have been conducted. As we will show below, besides bulk components,
peak iii) also contains components from the surface layer below 7 K.

Crucially, we also observe small anomalies iv) and v) at $\pm$3 meV, which
reside just inside the bulk gap $\Delta_2$ ({\it cf.} also Fig.\ \ref{Fig4}c
and d). The shoulder-like shape of these small anomalies indicates the
existence of two weakly dispersive bands or localized states near $E_F$.
It is noted that both features at about $\pm$3 meV also reveal spatial
inhomogeneity (see Supplementary Fig.\ 2), which---given the
electronic inhomogeneity of even atomically flat surfaces
\cite{roe15}---hints at a surface origin of these states.

\textbf{Spatial dependence of the STS spectra.}
For STM measurements, one possible way to distinguish
bulk and surface states is to carefully investigate
the tunneling spectra at/near impurities or other defects,
because the surface states are more vulnerable to such defects.
\begin{figure*}[t]
\centering\includegraphics[width=15cm]{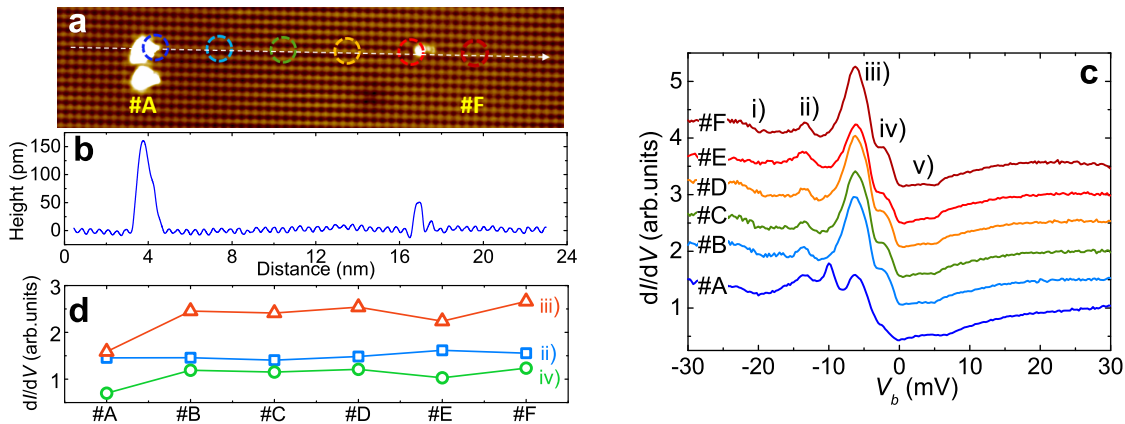}
\caption{\textbf{Spatial dependence of tunneling spectroscopy.}
(\textbf{a}) Topography of a non-reconstructed, B-terminated
surface (area of 24 $\times$ 4 nm$^2$) with two different
types of defects, two large ones at position $\#$A and a smaller
one at $\#$E. (\textbf{b}) Height scan along the dashed line indicated
in (\textbf{a}). (\textbf{c}) d$I$/d$V$-curves measured at six positions
(denoted as $\#$A to $\#$F) equally spaced and marked by circles
in (\textbf{a}). Curves are offset for clarity. $T$ = 0.35 K, $V_b$ = 30
mV, $I_{sp}$ = 100 pA, $V_{mod}$ = 0.3 mV. (\textbf{d}) Maximum peak
values of the differential conductance at $-13.5$ mV, $-6.5$ mV
and $-3$ mV obtained at positions $\#$A to $\#$F.} \label{Fig3}
\end{figure*}
Therefore, $g(V)$ was measured across two impurity sites at 0.35 K, shown
in Figs.\ \ref{Fig3}a and b. The bigger impurity at $\#$A
with an apparent height of $\approx$160 pm is probably located
on top of the surface, while the smaller one at $\#$E (apparent height
$\approx$50 pm) is likely incorporated into the crystal.
According to Fig.\ \ref{Fig3}c, the $g(V)$-curves are all very
similar for positions $\#$B to $\#$F. Even at position $\#$A,
{\it i.e.} on top of the big impurity, the spectrum exhibits
similarities; in particular all anomalies i)--v) can
be recognized. In addition, a new peak occurs at $-10$ meV,
which may be assigned to an impurity bound state. In Fig.\ \ref{Fig3}d,
we plot the height of the peaks ii) to iv) at different positions.
A combined analysis of Figs.\ \ref{Fig3}c and d reveals spatial
stability of peak ii), being consistent with the expectation for bulk
states as discussed above. On the other hand, peaks iii) and iv) are not
as stable as peak ii); their heights are suppressed by both the big and
the small impurity, which implies that at this temperature both peaks
contain contributions from the states pertained to the surface.

\textbf{Magnetic field dependence of the STS spectra.}
In Fig.\ \ref{Fig4}a and b, $g(V)$-curves measured at sites $\#$A and
$\#$C of Fig.\ \ref{Fig3}a for different applied magnetic fields are
presented. There is no distinct change detected up to the maximum field
of 12 T for features i) to v), except an enhanced peak amplitude for
the impurity state at $-10$ meV, see Fig.\ \ref{Fig4}b. The
magnetic-field independence of these states is consistent with the
observation of metallic surface conductance up to 100 T by transport
\cite{Chen,Biswas,Coo95b,Coo99} and spectroscopic measurements
\cite{Fla2001,Yee}. This observation can be understood by considering
a very small $g$-factor (0.1--0.2) of the $f$-electrons \cite{Ert2015}.

\textbf{Temperature dependence of the STS spectra.}
We now turn to the temperature dependence of the features i) to v).
The temperature evolution of the STS spectra was measured continuously
on the same unreconstructed, B-terminated surfaces away from any
defect between 0.35 and 20 K, see Fig. \ref{Fig4}c. Above 15 K, the
spectra show a typical asymmetric lineshape which arises from the
Fano effect \cite{mal09,Figgins}, being in good agreement with previous
work \cite{Sahana} (see Supplementary Fig.~4). Upon cooling,
the amplitude of peak iii) increases sharply, accompanied by a sudden
appearance of peaks iv) and v) below 7 K, with the latter effect being
beyond thermal smearing (see Supplementary Fig. 3). The low-temperature
evolution of the spectra is clearly seen after the measured
$g(V,T)$-curves were subtracted by the data at 20 K, see Fig. 4d.
In an effort to quantitatively investigate the evolution of the spectra
with temperature, we describe the low-temperature $g(V)$-curves by a
superposition of four Gaussian peaks on top of a co-tunneling model
(see Supplementary Fig. 4). However, fits to data obtained at higher
temperature ($T$ $>$ 10 K) turned out to be less reliable (Supplementary
Figs. 5 and 6).

To further analyze the temperature evolution of peak iii),
we normalized the spectra by its size at $V_b$ = $\pm$30 mV.
The resulting $g(T)$-values of peak iii)
are plotted in Fig.\ \ref{Fig4}e. Clearly, a change in the
temperature dependence is observed around 7 K. This is
further supported by a comparison to data obtained by Yee
{\it et al.} \cite{Yee} (blue circles and blue dashed line) in
a similar fashion but on a ($2\!\times \! 1$) reconstructed
surface (which may explain the scaling factor, right axis).
Also, the spectral weights of the $-10$ meV peak by Ruan
{\it et al.} \cite{Ruan} (green squares) indicate a similar
trend at $T$ $\geq$ 5 K. Note that even the temperature evolution
above about 7 K cannot be explained by a mere thermal broadening
effect \cite{Yee,Ruan}. By tracing the temperature evolution of
the d$I$/d$V$-spectra between about 7--50 K \cite{Yee,Sahana,Ruan}
a characteristic energy scale of about 50 K was derived. This can be
accounted for by the Kondo effect of the bulk states, with an
additional contribution from a resonance mode \cite{Ruan} which is
likely (as discussed above) related to the $\Gamma_8^{(1)}$ state.
The same energy scale of $\gtrsim 50$ K has also been observed by
transport \cite{All1979,Wol2013} and other spectroscopic
measurements \cite{Zhang,Xu2014,Caldwell,CHMin}.
However, below 7 K, the intensity of peak iii) shows
a sudden enhancement in Fig.\ \ref{Fig4}e,
indicating the emergence of an additional energy scale.
Considering the fact that this new energy scale as well as many other
exotic transport phenomena related to the formation of a metallic
surface \cite{Kim,Kim2014,Wol2013} set in simultaneously,
the increase in intensity of peak iii) (as well as the appearance of peaks iv)
and v)) is expected to rely on the same mechanism that is responsible for the
formation of the metallic surface. Both observations appear to evolve out of
the bulk phenomena associated with the primary hybridization gap
at elevated temperatures. In the following section, we will argue that this
new energy scale is related to the breakdown of the Kondo effect at the
surface.

\section*{Discussion}
In this study, the topographic capabilities of the STM allow us to
distinguish features i) to v) on non-reconstructed (001) surfaces
of a single termination and without apparent defects. Therefore, we
can simply exclude the possibility that they are driven by surface
reconstructions or defects. Especially the observation of new states on clean
surfaces below about 7 K indicates that the exotic properties of SmB$_6$ are
intrinsic rather than due to impurities. The observation of well-resolved
\begin{figure*}[t]
\centering\includegraphics[width=17cm]{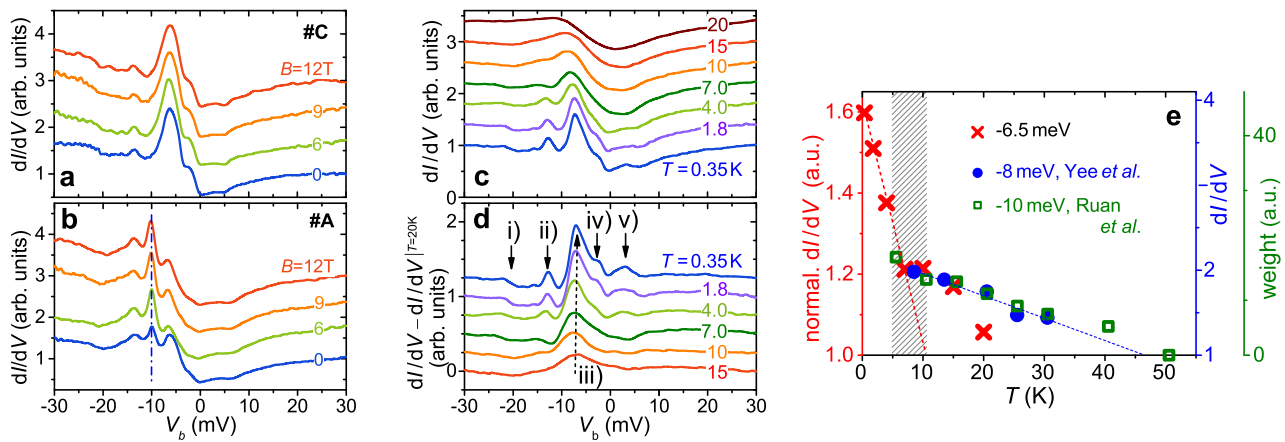}
\caption{\textbf{Magnetic field and temperature dependence of the STS spectra}.
Tunneling spectra measured at magnetic fields up to 12 T and 0.35 K on the
top of site (\textbf{a}) $\#$C and (\textbf{b}) $\#$A ({\it cf.} Fig.\
\ref{Fig3}\textbf{a}). (\textbf{c}) Evolution of d$I$/d$V$-curves from 0.35 K
to 20 K. In order to compare the data, a small linear background is
subtract from the raw data. (\textbf{d}) Difference tunneling conductance
after subtracting the $g(V)$-data measured at 20 K. (\textbf{e}) Temperature
dependence of the intensity of peak iii). Results for intensities by
Yee {\it et al.} \cite{Yee} (blue) and spectral weights by Ruan {\it et al.}
\cite{Ruan} (green) are shown for comparison (note matching colors of
markers and right axes). Curves are offset for clarity in
(\textbf{a})--(\textbf{d}). $V_b$ = 30 mV, $I_{sp}$ = 100 pA, $V_{mod}$ =
0.3 mV.}
\label{Fig4} \end{figure*}

features in our tunneling spectra (discussed above) enables the direct
comparison with results of bulk band structure calculations \cite{Lu2013,Denlinger,Junwon,Baruselli,Peters,KangJPSJ}.
This not only reveals the energy levels of the multiplet $f$-states, but can
also reconcile the long-standing debate of `small' versus `large' bulk gap
in SmB$_6$ \cite{Den2014}. Consequently, our data shows that a dedicated
hybridization model with two---instead of one---multiplet $f$-states is
necessary to interpret the low-energy properties of SmB$_6$. In particular,
peak iii) has multiple components including bulk and surface states, the
ratio of which changes dramatically with temperature.

It is widely accepted that the electronic properties of SmB$_6$ can
be divided into several temperature regions, which are based on transport
measurements \cite{Denlinger,Chen} as well as other probes,
like ARPES \cite{Denlinger,Xu2014}. Apparently, 5-7 K is a crucial regime
where the temperature-dependent properties undergo significant changes. Above
this range, the electronic states in SmB$_6$ are
governed by the Kondo effect of the bulk \cite{Den2014,Jiang,Frantzeskakis}.
At lower temperatures, several interesting observations---in addition to that
of the saturated resistance---were made. For example, the
Hall voltage becomes sample-thickness independent \cite{Kim}; the
angular-dependent magnetoresistance pattern changes from fourfold to
twofold symmetry \cite{Chen}; and the development of a heavy fermion surface
state is found by magnetothermoelectric measurements \cite{Luo}. These
experimental facts provide convincing evidence for the formation of (heavy)
surface states just around 5-7 K, which is in line with the
appearance of a new energy scale.

Recently, a surface Kondo breakdown scenario was proposed based on
the reduced screening of the local moments at the surface. As a result,
the Kondo temperature of the outmost layer ($T_K^s$) can be strongly
suppressed, resulting in a modified band structure \cite{Alexandrov}.
Slab calculations further show that below $T_K^s$ $f$-electrons
gradually become coherent and form weakly dispersive band
close to $E_F$ \cite{Ert2015,Peters}. Remarkably, very narrow peaks with
strongly temperature-dependent STS spectra near $E_F$ are regarded as a
smoking gun evidence for a surface Kondo breakdown scenario \cite{Peters}.
Based on our experimental results, $T_K^s$ is inferred to be around 7 K,
being about an order of magnitude smaller than $T_K$. The evolution of
our tunneling spectra below about 7 K also fit excellently to the
theoretical prediction and the related calculations for STS. In such a scenario,
the additional component at $-6.5$ meV and shoulders at $\pm$3 meV are related
to the heavy quasiparticle \emph{surface} states, the formation of which
supplies an additional tunneling channel in particular into the $f$-states.
This provides a highly possible origin for the metallic surface states and a
reasonable explanation to the various experimental observations
listed above.

We note that theoretically a surface Kondo breakdown effect does not 
change the topological invariance of SmB$_6$, which is determined by the
topology of the bulk wave functions. Therefore, the surface-derived heavy
quasiparticle states could still be topologically protected. Experimentally,
for such topologically protected surface states backscattering is forbidden 
in quasiparticle interference (QPI) patterns as measured by STM \cite{Xu2016}.
In line with this prediction and as shown in the Supplementary Fig. 7, no 
clear QPI pattern could be detected so far from our results, which is similar
to the observation by Ruan {\it et al.}, \cite{Ruan}.

\section*{Methods}
All samples were grown by the Al-flux method. A cryogenic (base temperature 
$T \approx 0.35$ K) STM with magnetic field capability of $\mu_0 H \leq 12$
T was utilized. Three SmB$_6$ single crystals were cleaved a total of five
times \textit{in situ} at $\approx$20 K to expose a (001) surface. Cleaved
surfaces were constantly kept in ultra-high vacuum, $p < 3 \cdot 10^{-9}$ Pa.
Tunneling was conducted using tungsten tips \cite{Ernst} and the differential
conductance ($g(V)$-curve) is acquired by the standard lock-in technique with
a small modulation voltage. In our best cleaved sample, the size of
non-reconstructed surface can reach to $100\, \times\, 100$ nm$^2$.

In principle, the low-temperature $g(V)$-curves can be well described by a
superposition of four Gaussian peaks on top of a Fano model (see example of
$g(V,T\! =\! 0.35\,$K$)$ in Supplementary Fig.~4) or more elaborate
hybridization models \cite{mal09,Figgins} (Supplementary Fig.~6). A similar
procedure with only one Gaussian was employed in \cite{Ruan}. However, fits
are less reliable at elevated temperature. Instead, our spectra measured at
different $T$ in zero field overlap nicely for $V_b < -25$ mV and $V_b > 10$
mV such that they can be normalized by using very similar factors.
Consequently, we can directly trace the temperature dependence of the peak
height (at least for peak iii)) by measuring the normalized peak intensity 
as shown in Fig.\ \ref{Fig4}e. Note that the intensities of peak iii) as
obtained from Fig.\ \ref{Fig4}d, {\it i.e.\ }after subtracting the 20 K-data,
yield very similar values as those shown in Fig.\ \ref{Fig4}e from normalized
spectra.

\section*{acknowledgments}
We acknowledge valuable discussion with J.~W. Allen, P.
Coleman, X. Dai, I. Eremin, O. Erten, C.-L. Huang, Deepa Kasinathan,
B.~I. Min, C.~J. Kang, G.~Sawatzky, Q. Si and P.~Thalmeier.
This work was supported by the Deutsche Forschungsgemeinschaft
through SPP 1666 and by the Defense Advanced Research Agency
(DARPA) under agreement number FA8650-13-1-7374. L.J.
acknowledges support by the Alexander-von-Humboldt foundation.

\end{document}